# Pressure-Driven Quantum Criticality in Iron-Selenide Superconductors

Jing Guo[1]*, Xiao-Jia Chen[2,3]*, Jianhui Dai[4], Chao Zhang[1], Jiangang Guo[1], Xiaolong Chen[1], Qi Wu[1], Dachuan Gu[1], Peiwen Gao[1], Lihong Yang[1], Ke Yang[5], Xi Dai[1], Ho-kwang Mao[2], Liling Sun[1]†, and Zhongxian Zhao[1]††

[1] Institute of Physics and Beijing National Laboratory for Condensed Matter Physics, Chinese Academy of Sciences, Beijing 100190, China
[2] Geophysical Laboratory, Carnegie Institution of Washington, Washington, D.C. 20015, USA
[3] Department of Physics, South China University of Technology, Guangzhou 510640, China
[4] Department of Physics, Hangzhou Normal University, Hangzhou 310036, China
[5] Shanghai Synchrotron Radiation Facilities, Shanghai Institute of Applied Physics, Chinese Academy of Sciences, Shanghai 201204, China

We report a finding of pressure-induced quantum critical transition in $K_{0.8}Fe_xSe_2$ (x=1.7 and 1.78) superconductors through *in-situ* high-pressure electrical transport and X-ray diffraction measurements in diamond anvil cells. Transitions from metallic Fermi liquid behavior to non-Fermi liquid behavior and from antiferromagnetism to paramagnetism are found in the pressure range of 9.2-10.3 GPa, in which superconductivity tends to disappear. The change around the quantum critical point from the coexisted antiferromagnetism state and the Fermi liquid behavior to the paramagnetism state and the non-Fermi liquid behavior in the iron selenide superconductors demonstrates a unique mechanism for their quantum critical transition.

The recent discovery of superconductivity in $K_{0.8}Fe_2Se_2$ selenide [1] with a transition temperature ($Tc$) above 30 K has generated considerable interest because its isostructure $KFe_2As_2$ pnictide only has a $Tc$ of about 3 K and the selenide is more environmentally friendly than the pnictide. Later, superconductivity in other $A_xFe_{2-y}Se_2$ (A =Rb, Cs, or Tl substituted K, Rb) compounds has been also found [2–4]. The carriers in these superconductors were identified to be electrons from the measurements of optical spectroscopy [5], Hall effect [6] and angle-resolved photoemission spectroscopy [7–9]. This is quite different from pnictide superconductors which have both electron and hole pockets at the Fermi surface [10]. Superconductivity of such iron selenides was reported to coexist with antiferromagnetism (AFM), its ordering temperatures $T_N$ as high as ~550 K [11–15], and large magnetic moments of 3.3 $\mu_B$ for each Fe atom [11, 15]. Theoretical [16] and experimental [11] studies on $K_{0.8}Fe_{1.6}Se_2$, which is thought to be the parent compound of these superconductors, revealed that the ground state of this compound is in reality a quasi-two-dimensional blocked checkerboard antiferromagnetic semiconductor (or insulator). The Fe vacancies have been proposed to be the major players of the observed superconductivity and many interesting physical properties [15, 17-18].

Superconductivity has been thought to be closely related to the quantum critical transition (QCT) in many correlated electronic systems such as cuprates [19-22], heavy fermions [23-24], organic conductors [25-26] and iron pnictides [27-29]. The quantum states are determined by the lattice, charge, orbital and spin degrees of freedom in materials. These factors can be manipulated by control parameters including pressure, magnetic field and chemical composition. Among these parameters, pressure is a clean way in tuning lattice and electronic structures, as well as the interaction between them.

In this Letter, we report an experimental discovery of pressure-driven quantum criticality in the newly discovered iron-selenide superconductors $K_{0.8}Fe_xSe_2$ (x=1.7 and 1.78) through resistance and structure measurements. We find that superconductivity in the two superconductors investigated is gradually suppressed with the applied pressure and eventually disappears at 9.2 GPa for $K_{0.8}Fe_{1.7}Se_2$ and 9.7 GPa for $K_{0.8}Fe_{1.78}Se_2$, around the pressure of which the superstructure of Fe vacancies vanishes and the metallic non-Fermi liquid (NFL) behavior characterized by linear-temperature-dependent resistance persists over a wide temperature region. Meanwhile, the activation energy for the electronic transport of the high-temperature resistance approaches to zero. The presence of such pressure-induced QCT classifies the iron-selenide superconductors into the quantum matter with quantum criticality.

High-pressure electrical resistance measurements on $K_{0.8}Fe_xSe_2$ (x=1.7 and 1.78) single crystals were carried out in a diamond anvil cell made from Be-Cu alloy in a house built refrigerator. The chemical composition was identified by using ICP analysis. Diamond anvils of 600 and 300 μm flats were used and the corresponding sample holes with 300 μm and 100 μm in diameter were made in rhenium gaskets for the individual two runs. Insulation from the rhenium gasket was achieved by a thin layered mixture of c-BN (cubic boron nitride) powder and epoxy. The crystal was placed on the top anvil and then pressed into the insulating gasket hole with leads. NaCl powders were employed as pressure medium. The pressure was determined by the ruby fluorescence method [30]. A standard four-probe technique was adopted in these measurements. Electrical resistance measurements at ambient pressure and magnetic field were performed using a Quantum Design Physical Property Measurement System. Powder X-ray diffractions performed at ambient and high pressure were used to obtain the structural information based on the powders from the cleaved pieces of crystals. Rietveld refinements were performed by using the FULLPROF package [31].

Figure 1(a) shows the temperature dependence of the in-plane resistance of a $K_{0.8}Fe_{1.7}Se_2$ single crystal sample at various pressures. The superconducting transition occurs at 32.5 K and reaches zero resistance at 30.6 K at ambient pressure. A remarkable feature of this superconductor is that its resistance exhibits a large hump showing a crossover from semiconducting behavior to metallic behavior at $T_H$. This hump phenomenon has not been found in FeAs-based superconductors whose normal resistance behavior is metallic. Interestingly, the maximum resistance at $T_H$ is dramatically reduced when pressure is applied. Simultaneously, $Tc$ is suppressed and disappears at pressure above 9.2 GPa (Fig. 1(b)). Releasing pressure from 9.2 GPa, both $Tc$ and the resistive hump are recovered together, strongly suggesting that both phenomena are interconnected. We also performed high-pressure resistance measurements for $K_{0.8}Fe_{1.78}Se_2$ single crystal, whose composition is slightly different from $K_{0.8}Fe_{1.7}Se_2$, and found the same behavior of pressure-induced suppression of superconductivity in this compound, indicating that it is common that pressure has a negative effect on the superconductivity of this kind of iron-selenide superconductors.

Figure 2a shows the pressure dependence of $Tc$ of the two samples with x=1.7 and 1.78. $Tc$ exhibits a systematic reduction with pressure and disappears at the lowest temperature 4.2 K of our refrigerator above 9.2 GPa. To understand the picture emerged in the present study, we performed high-pressure X-ray diffraction measurement at beamline BL15U1 of the Shanghai Synchrotron Radiation Source for the sample $K_{0.8}Fe_{1.78}Se_2$. The most striking feature of the studied compounds is the existence of the correlation between the Fe-vacancies ordered with a √5×√5 superstructure in the Fe-square lattice and an unusual AFM order with large magnetic moment per Fe atom [11]. From the experimental results reported [11, 14], it is found that the ordered Fe-vacancies construct the AFM order in the iron selenide superconductors investigated. Once the Fe-vacancy ordering is absent, its superstructure peak is disappeared and the sample undergoes a transition from a AFM state to a paramagnetic (PM) state [11, 14], which allows us to trace the magnetic structure evolution with pressure by the way of characterization of Fe's superstructure peak. As seen in Fig. 2b, we found that a tetragonal phase with *I4/m* symmetry exists in the sample at pressure below 9.2 GPa. With increasing pressure to 10.3 GPa, the superstructure peak (110) is completely suppressed, revealing the full suppression of AFM ordering with applied pressure. It is worthy to note that superconductivity vanishes when the superstructure peak of the Fe-vacancy ordering is absent. The results give an evidence for the presence of pressure-induced QCT in the iron selenide superconductor. We propose that the quantum critical point (QCP) should exist in the pressure range between 9.2 and 10.3 GPa. Below the QCP, the sample studied is in the AMF state with *I4/m* symmetry, while above the QCP, the sample has *I4/mmm* symmetry and loses its AFM ordering.

Pressure-driven magnetic transition in this kind of superconductor has been observed recently through $^{57}$Fe-Mössbauer measurements [32]. It has been shown a clear AFM-to-PM phase transition in the compressed superconductor. At transition pressure, its superconductivity is completely suppressed. This suggests that application of pressure can suppress the AFM long-range ordering and produce a new magnetic state.

We also make the actual fits to the temperature dependence of the normal state resistance in low temperature side (slightly below $T_H$) for the data obtained at each pressure point, based on the form of $\rho=\rho_0+AT^\alpha$. We found that the power $\alpha$ is pressure dependent, varying from initial $\alpha$=2.7 at ambient pressure for both superconductors to $\alpha$=1 at pressure above 9.2 GPa for $K_{0.8}Fe_{1.7}Se_2$ and above 9.7 GPa for $K_{0.8}Fe_{1.78}Se_2$, as shown in Fig.2a. The electron response of the systems with pressure suggests that application of pressure drives the systems undergoing a transition from a Fermi liquid (FL) behavior to a NFL behavior, accompanying the transition from superconducting to non-superconducting state. These findings provide further evidence for the existence of the pressure-induced quantum critically in $Ke_{0.8}Fe_xSe_2$ (x=1.7 and 1.78) superconductors.

We used the structural determination to clarify the origin of the resistance hump at $T_H$. Figure 3(a) shows the ambient pressure X-ray diffraction patterns at selected temperatures down to 60 K. The data demonstrate that the sample has a tetragonal ThCr$_2$Si$_2$-type structure with space group *I4/m* over the temperature range crossing $T_H$, indicating that no structural transition can be detected. As expected, the lattice parameters *a* and *c* decrease smoothly with decreasing temperature (Figs. 3(b) and 3(c)). Our data offer clear evidence in supporting that the hump is irrelevant to any structural transition.

Figure 4 shows the temperature dependence of the resistance of K$_{0.8}$Fe$_{1.7}$Se$_2$ at magnetic fields of 0, 3 and 7 T. As seen, $T_H$ is nearly unchanged when the magnetic field is applied. Since the hump feature is neither related to a magnetic transition nor a structural transition, we propose that this feature may result from a competition between semiconducting state and metallic state in the superconducting sample, in which semiconducting behavior is dominated at higher temperature region above $T_H$ but metallic behavior is prevailed below $T_H$.

Remarkable increase in $T_H$ is observed in the sample at high pressure as shown in Fig. 5a. The phase transformation occurred above $T_H$ from low-pressure semiconductor to the high-pressure metal takes place at around 9 GPa. This is the pressure at which superconductivity disappears, as shown in Fig. 2. The exact transition pressure is estimated from the pressure dependence of the activation energy for the electrical transport in the high-temperature semiconducting state (Fig. 5(b)). The extension of the fitting curve yields a critical pressure of 8.7 GPa. Here, the activation energy is obtained by fitting the temperature dependence of the resistance in terms of an Arrhenius equation. The reduction of activation energy ($E_A$) with the applied pressure is suggested to originate from the mechanism: pressure minimizes the gap which gives rise to a semiconducting-to-metal transition. As a result, the remarkable increase in $T_H$ with pressure can be understood by the scenario that the pressure-induced gap shrinkage enhances the metallicity of the sample.

The transition from the AFM phase to the PM phase determined by characterization of superstructure of ordered Fe vacancies and the transition from a metallic FL behavior to a NFL behavior, together with the phase transformation above $T_H$ from the low-pressure semiconductor to high-pressure metal and from superconducting phase to non-superconducting phase, demonstrate that a QCP exists at ~10 GPa . We noted that the change around the QCP from the coexisted AFM state and FL behavior to the PM state and NFL behavior in these iron selenide superconductors is different from that of copper oxide superconductors [22]. In this strongly correlated electronic oxide materials, the FL behavior appears in the overdoped system in which the AFM ordering no longer exists and superconductivity is absent [22, 33-34]. Therefore, the very large Fe moment [11, 14] and the unusual FL behavior in this kind of iron selenide superconductor indicate that its quantum criticality is quite unique.

It is generally believed that a QCT in an electron system can induce a quantum correlated state, from which a superconducting state emerges below a certain temperature, resembling that seen in cuprates and heavy fermion superconductors [22, 24, 33]. In the extended high-pressure studies on these iron selenide superconductors, we found a new superconducting phase appears at pressure around 10.5 GPa for $K_{0.8}Fe_{1.7}Se_2$ and 10.7 GPa for $K_{0.8}Fe_{1.78}Se_2$ single crystals after the elimination of the initial superconducting phase [35]. The maximum *Tc* of the second superconducting phase reached 48 K, higher than the maximum *Tc* of the first superconducting phase. From the results of this Letter, we note that the emergence of the second superconducting phase is at the pressure where the system loses its AFM ordering. Accordingly, we propose that the reemerging superconductivity in the studied samples should be driven by the quantum critically reported in this Letter.

In summary, we reported the pressure-driven quantum criticality in the newly discovered superconductors $K_{0.8}Fe_xSe_2$ (x=1.7 and 1.78) through a systematic investigation of electrical transport and structural properties. Upon approaching the QCP around 10 GPa, superconductivity tends to disappear and the activation energy for the electrical transport of the high-temperature resistance goes to zero. We have presented experimental evidence for the coexistence of AFM state and FL behavior below the QCP, as well as the coexisted PM state and NFL behavior above the QCP. The observed quantum criticality may provide important information in shedding the insight on the underlying mechanism of superconductivity in the iron-selenide superconductors.

We thank I. I. Mazin and F. Steglich for reading this paper and T. Xiang, Z. Fang and G. M. Zhang for valuable discussions. This work was supported by the NSCF (10874230, 10874046, and 11074294), 973 projects (2010CB923000 and 2011CBA00109), and Chinese Academy of Sciences. Work done in the USA was supported as part of the EFree, an Energy Frontier Research Center funded by the US Department of Energy, Office of Science, Office of Basic Energy Sciences (DOE-BES) under Grant No. DE-SC0001057.

*These authors contributed equally to this work.

†‡ Corresponding authors: llsun@aphy.iphy.ac.cn and zhxzhao@aphy.iphy.ac.cn

**Figure captions:**

Figure 1 (Color online) Temperature dependence of the electrical resistance of a $K_{0.8}Fe_{1.7}Se_2$ single crystal measured at different pressures and in the temperature range of 4.2-290 K (a) and of 4.2-50 K (b). The arrow in (a) shows a transition temperature $T_H$ of the resistance in a hump shape from its high-temperature semiconducting to low-temperature metallic behavior. The arrow in (b) denotes the superconducting transition temperature $Tc$.

Figure 2 (Color online) (a) Pressure dependence of the superconducting transition temperature $Tc$ and power α obtained from fits with $\rho=\rho_0+AT^\alpha$ for $K_{0.8}Fe_xSe_2$ (x=1.7 and 1.78) single crystals. (b) The X-ray diffraction patterns of $K_{0.8}Fe_{1.78}Se_2$, performed with a wavelength of 0.6888 Angstrom. (c) Intensity of superstructure peak (110) of Fe vacancies as a function of pressure. The inset of Fig.2c display the schematics of the AFM state, in which the spin order is ferromagnetic groups oriented along the *c*-axis and couples antiferromagnetically, and the PM state in higher pressure region.

Figure 3 (Color online) (a) X-ray diffraction patterns of the $K_{0.8}Fe_{1.7}Se_2$ sample collected at different temperatures and ambient pressure. (b) and (c) The refined lattice parameters *a* and *c* as a function of temperature.

Figure 4 (Color online) Temperature dependence of the resistance of a $K_{0.8}Fe_{1.7}Se_2$ single crystal with magnetic fields applied along the *c* axis of 0, 3, and 7 T, respectively. The inset is an extended view with normalized resistance at 35 K around the superconducting transition.

Figure 5 (Color online) Pressure dependence of (a) $T_H$ in the hump shape and (b) the activation energy of the electric transport of the high-temperature resistivity of a $K_{0.8}Fe_{1.7}Se_2$ single crystal. The line in (b) is the linear fitting to the data points. The vertical dashed line denotes the phase boundary.

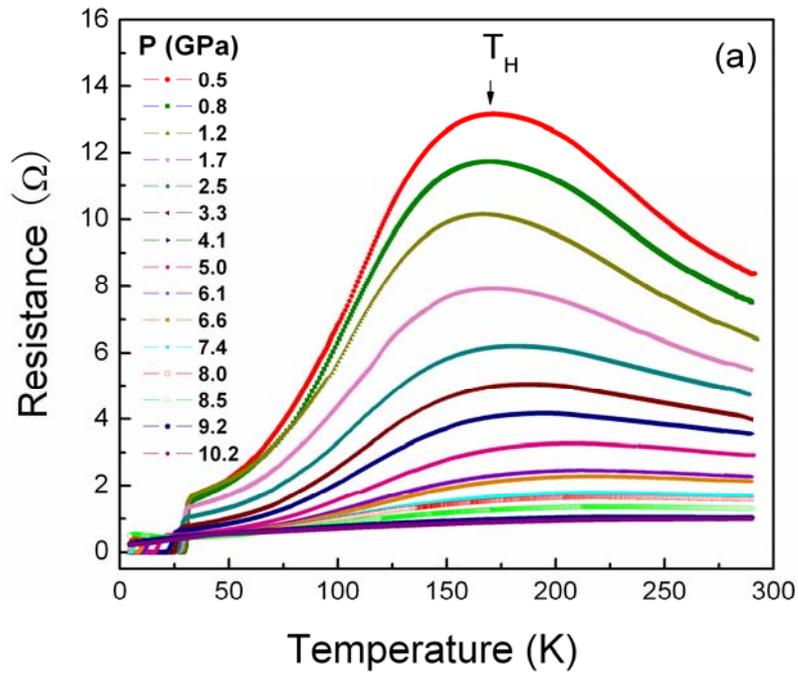

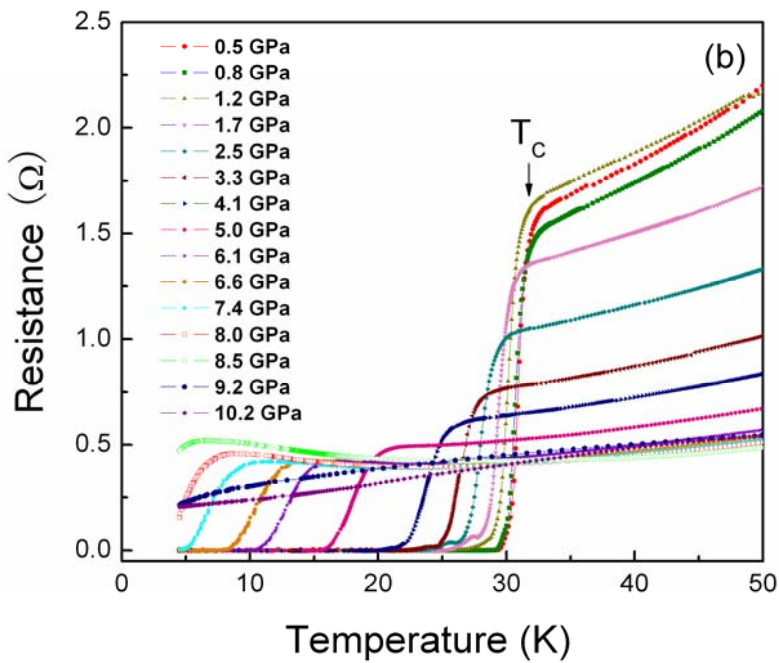

Fig.1 Guo et al

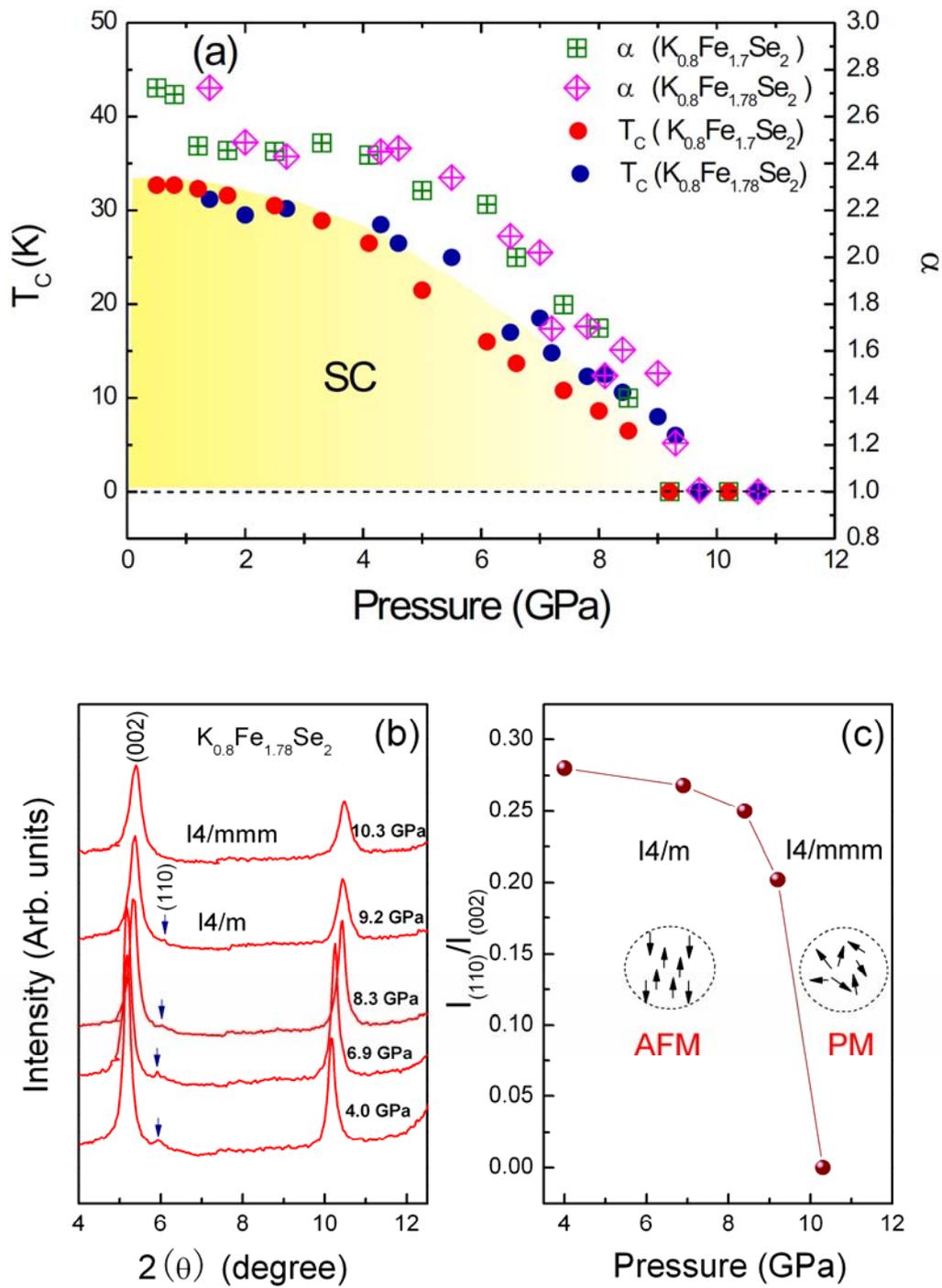

Fig.2 Guo et al

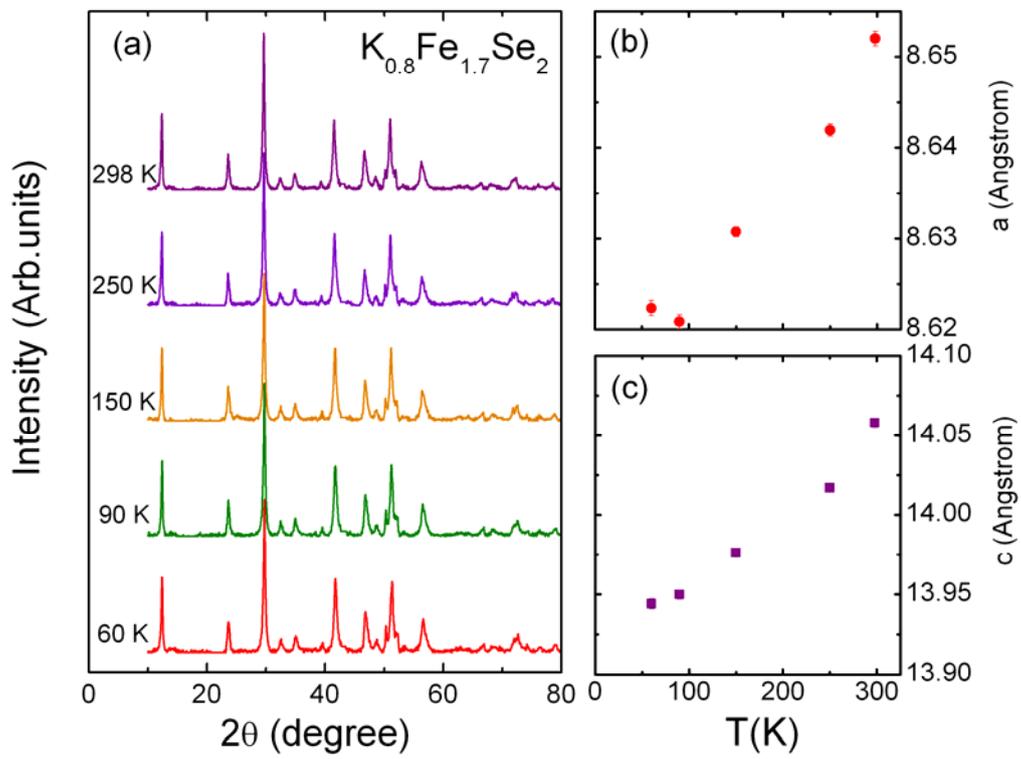

Fig. 3 Guo et al

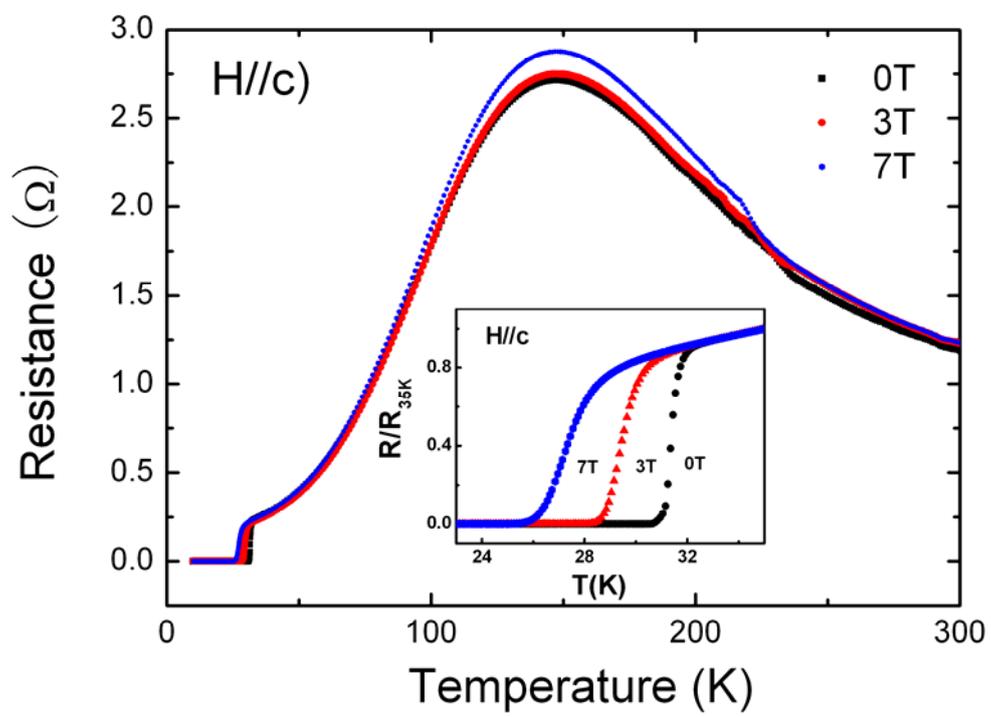

Fig. 4 Guo et al

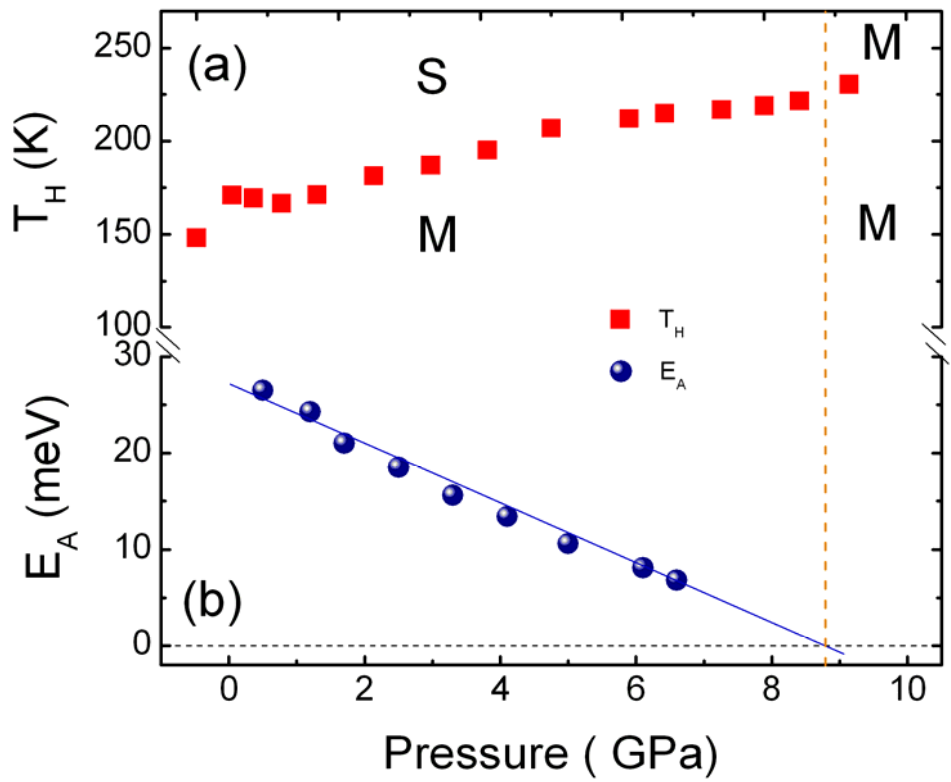

Fig. 5 Guo et al